\documentstyle[tighten,preprint,aps,overcite]{revtex}

\begin{document}

\title{GRECP/RCC calculation of
the spectroscopic constants for the HgH molecule and its cation}
\author{
Nikolai S. Mosyagin,\thanks{
  E-mail for correspondence: Mosyagin@lnpi.spb.su;
  http://www.qchem.pnpi.spb.ru}
        Anatoly V. Titov}
\address{
 Petersburg Nuclear Physics Institute, \\
         Gatchina, St.-Petersburg district 188300, Russia}
\author{Ephraim Eliav, Uzi Kaldor}
\address{
School of Chemistry, Tel Aviv University,
         Tel Aviv 69978, Israel}
\maketitle

\begin{abstract}
Generalized relativistic effective core potential (GRECP) calculations of
spectroscopic constants of the HgH molecule ground and low excited states
and the HgH$^+$ cation ground state are carried out,
with correlation included by the Fock-space relativistic coupled cluster (RCC)
method. Basis set superposition errors (BSSE) are estimated and discussed.  It
is demonstrated that connected triple excitations of the 13 outermost
electrons are necessary to obtain accurate results for mercury hydride.
Spectroscopic constants derived from potential curves which include these
terms are in very good agreement with
experiment, with errors  of a few mbohr in $R_e$, tens of
wavenumbers in excitation energies and vibrational frequencies, and
proportionately for other properties.
Comparison with previous calculations is also presented.
\end{abstract}


\section{Introduction}

The HgH molecule has been studied in the last few decades both experimentally
(see, e.g., references \citen{Porter,Stwalley,Mayama,Dufayard}) and
theoretically (references \citen{Das,Hausser,Alekseyev,Motegi}, among others).
The calculations used 12-, 18- or 20-electron relativistic effective core
potentials (RECPs), all-electron quasirelativistic methods,
and a variety of approaches accounting for electron correlation.
The goals of the theoretical investigations included the explanation and
systematization of available experimental data, as well as assessing the
accuracy and reliability of methods for calculating molecules containing heavy
elements.

A 20-electron generalized relativistic effective core potential (20e-GRECP)
has been generated for mercury~\cite{Tupitsyn,Mosyagin}
and tested in numerical two-component Hartree-Fock (HF)
calculations by comparison with other RECP and all-electron Dirac-Fock (DF)
studies. The suitability of the GRECP for describing
correlation effects was examined in atomic calculations.\cite{Mosyagin1}
Significant improvement in the accuracy of reproducing all-electron
Dirac-Coulomb data for the GRECP as compared with other RECPs\cite{Hausser,Ross}
treating explicitly the same number of electrons was demonstrated in these
calculations. Here we present GRECP calculations of
spectroscopic constants for the HgH molecule and its cation in the framework
of the relativistic coupled cluster (RCC) method.

\section{The GRECP method}

The GRECP method has been described in detail
elsewhere\@.\cite{Tupitsyn,Mosyagin,Titov} In this method, the radial
oscillations of the valence and outer core spinors are smoothed in
the inner core region of an atom to reduce the number of primitive
Gaussian basis functions required for
appropriate description of these spinors in subsequent molecular
calculations.  This smoothing also allows one to exclude the small
components of the four-component Dirac spinors from the GRECP calculations,
with relativistic effects being taken into account by
{\it j}-dependent effective potentials. These $U_{nlj}$ potentials are
derived by inversion of the nonrelativistic-type HF equations in the
$jj$-coupling scheme for a ``pseudoatom'', in which the inner core
electrons are removed:

\begin{eqnarray}
 U_{nlj}(r)  & = & \widetilde{\varphi}_{nlj}^{-1}(r)
                \biggl[\biggl( \frac{1}{2}
		{\bf \frac{d^{2}}{dr^{2}} }
                - \frac{l(l+1)}{2r^{2}}
                + \frac{Z_{ic}}{r} -
                \widetilde{\bf J}(r) +
                \widetilde{\bf K}(r)
                        \nonumber\\
             & +  &  \varepsilon_{nlj}  \biggr) \widetilde{\varphi}_{nlj}(r) +
                \sum_{n'\neq n} \varepsilon_{n'nlj}
                \widetilde{\varphi}_{n'lj}(r) \biggr],
\label{U_nlj}
\end{eqnarray}
where $Z_{ic}$ is the charge of the nucleus and inner core electrons,
$\widetilde{\bf J}$ and $\widetilde{\bf K}$ are Coulomb and
exchange operators calculated with the $\widetilde{\varphi}_{nlj}$
pseudospinors, $\varepsilon_{nlj}$ are the one-electron energies of the
corresponding spinors, and $\varepsilon_{n'nlj}$ are off-diagonal Lagrange
multipliers. The GRECP components, $U_{nlj}$, are usually fitted by Gaussian
functions to be employed in molecular calculations with Gaussian basis sets.
Potentials in conventional RECPs are constructed only for
nodeless pseudospinors, because division by zero appears in
equation~(\ref{U_nlj}) for pseudospinors with nodes. This problem is overcome
in the GRECP method by interpolating the
potentials in the vicinity of these nodes\@.\cite{Titov1}
This allows one to generate different potentials, $U_{clj}$ and $U_{vlj}$, for
outer core and valence pseudospinors, unlike the conventional RECP approach.

The GRECP operator has the form
\begin{eqnarray}
 {\bf U} & = & U_{vL}^{\rm AGREP}(r) + \sum_{l=0}^{L-1}
                   \Bigl[U_{vl}^{\rm AGREP}(r)-U_{vL}^{\rm AGREP}(r)\Bigr]
                   {\bf P}_{l}
                +  \sum_{l=1}^{L} {\bf U}_{vl}^{\rm EGSOP} {\bf P}_{l}
		   \nonumber\\
             & + & \sum_{c}\sum_{l=0}^{L} {\bf U}_{cl}^{\rm AGREP}(r)
                     {\bf P}_{l}+
                   \sum_{c}\sum_{l=1}^{L}{\bf U}_{cl}^{\rm EGSOP}
		     {\bf P}_{l},
 \label{GRECP_LS}
\end{eqnarray}
\[
   {\bf P}_{l} = \sum_{m=-l}^{l} | lm \rangle \langle  lm |,
\]
where $|lm\rangle \langle lm|$ is the projector on the spherical function
$Y_{lm}$ and $L$ is one more than the highest orbital angular momentum of
the inner core spinors.

 The ${\bf U}$ operator
 may be split into spin-averaged and spin-dependent parts.
 The first part
includes spin-averaged generalized relativistic effective
potentials (AGREP), given by
\begin{eqnarray}
   U_{vl}^{\rm AGREP}(r) & = & \frac{l+1}{2l+1} U_{vl+}(r)
		    + \frac{l}{2l+1} U_{vl-}(r), \nonumber\\
\label{AGREP}
   {\bf U}_{cl}^{\rm AGREP}(r) & = & \frac{l+1}{2l+1}
                      {\bf V}_{cvl+}(r)
                    + \frac{l}{2l+1} {\bf V}_{cvl-}(r),    \\
   {\bf V}_{cvl\pm}(r) & = & \bigl[ U_{cl\pm}(r)-U_{vl\pm}(r) \bigr]
                               \widetilde{{\bf P}}_{cl\pm}(r)
                               +\widetilde{{\bf P}}_{cl\pm}(r)
                               \bigl[U_{cl\pm}(r)-U_{vl\pm}(r)\bigr] \nonumber\\
                         & - & \sum_{c'} \widetilde{{\bf P}}_{cl\pm}(r)
                               \biggl[\frac{U_{cl\pm}(r)+U_{c'l\pm}(r)}{2}
                               -U_{vl\pm}(r)\biggr]
                                \widetilde{{\bf P}}_{c'l\pm}(r) , \nonumber
\end{eqnarray}
 where $\widetilde{{\bf P}}_{cl\pm}(r)$ is the radial projector on the
 outer core pseudospinor $\tilde{\varphi}_{cl\pm}(r)$ and $\pm$ means
 $j=|l\pm 1/2|$. Obviously, these components can be used to account for
 spin-independent relativistic effects in codes employing
 the $\Lambda S$-coupling scheme.
The other terms in ${\bf U}$ are
 the components of the effective spin-orbit interaction operator,
 called effective generalized spin-orbit potentials (EGSOP),
\begin{eqnarray}
   {\bf U}_{vl}^{\rm EGSOP} & = & \frac{2}{2l+1}[U_{vl+}(r) - U_{vl-}(r)]
   {\bf P}_{l} \vec{{\bf l}}\cdot\vec{{\bf s}}, \nonumber \\
   {\bf U}_{cl}^{\rm EGSOP} & = & \frac{2}{2l+1}[{\bf V}_{cvl+}(r) -
   {\bf V}_{cvl-}(r)]
   {\bf P}_{l} \vec{{\bf l}}\cdot\vec{{\bf s}}.
\label{EGSOP}
\end{eqnarray}

Two of the major features of the GRECP method are generation of effective
potential components for pseudospinors which may have nodes, and adding
non-local terms with projectors on the outer core pseudospinors
(second line in equation~(\ref{GRECP_LS})) to the standard semi-local terms
(first line in equation~(\ref{GRECP_LS})) of the effective potential
operator.
Other distinctive features of the GRECP generation
as compared to previous RECP schemes~\cite{Ross,Christ} are given
in reference \citen{Comment1}.
As pointed out earlier,\cite{Mosyagin,Titov}
the form~(\ref{GRECP_LS}) of the GRECP operator is optimal
for calculating states in which changes in occupation numbers
of outer core shells relative to the state used for the GRECP generation
are much smaller than 1.

\section{The RCC method}

The Fock-space RCC method has been described in previous papers\cite{Eliav}
and reviews,\cite{kaldorrevs} and only a brief summary is given here. Starting
from the nonrelativistic-type Hamiltonian ${\bf H}$ containing the AGREP
operator ${\bf U}^{\rm AGREP}$, the one-electron HF orbitals are obtained in
an SCF procedure.  Matrix elements of the EGSOP operator,~${\bf U}^{\rm
EGSOP}$, as well as other one- and two-electron integrals, are calculated in
the basis of the resulting spin-orbitals.  Spin-orbit interaction (described
by the ${\bf U}^{\rm EGSOP}$ operator) and correlation are then included by
the two-component Fock-space coupled-cluster (CC) method, using the
exponential valence-universal wave operator ${\bf \Omega} = \exp({\bf T})$.
The RCC expansion is currently truncated at the singles and doubles level
(RCC-SD).  The nonrelativistic CC (NRCC) version, in which the spin-orbit
interaction is not taken into account, allows one to include also the triple
cluster amplitudes (NRCC-SDT).  In the Fock-space CC method one starts from a
reference state (closed-shell in our implementation), correlates it, then adds
(or removes) one electron, recorrelating the new $N{+}1$ (or $N{-}1$) electron
system, and so on, until all the states of interest are attained. The
electrons can be added in (or removed from) a number of valence spin-orbitals,
resulting in a multireference approach characterized by a model space $P$ of
some selected states.  The cluster amplitudes are determined at the stage (or
Fock-space sector) where they first appear and are not changed at higher
sectors, leading to a valence-universal wave operator. The effective
Hamiltonian

\begin{equation}
{\bf H}_{\rm eff} = {\bf PH\Omega P} ,
\end{equation}
where ${\bf P}$ is the projector onto the model space $P$,
is diagonalized to give simultaneously the energies of all $P$-space
states correlated at the RCC-SD level relative to the correlated energy
of the reference state.

Two series of Fock-space calculations were performed. The ground state of
the HgH$^+$ ion served as reference in the first series (denoted RCC-SD-1),
and the Fock-space scheme was

\begin{equation}
 {\rm HgH}^+    \rightarrow  {\rm HgH} ,
\label{FS1}
\end{equation}
with electrons added in the lowest unoccupied $\sigma$ and $\pi$
orbitals of HgH$^+$.
The second series (RCC-SD-2) started from the ground state
of the HgH$^-$ ion as reference, using the Fock-space scheme
\begin{equation}
{\rm HgH}^- \rightarrow {\rm HgH} \rightarrow {\rm HgH}^+ ,
\label{FS2}
\end{equation}
with electrons removed from the highest occupied $\sigma$ orbital of HgH$^-$.

The MOLGEP~\cite{MOLGEP} and MOLCAS~\cite{MOLCAS} codes were employed in
the molecular GRECP/SCF calculations. The RCC-SD and NRCC-SDT
program packages were interfaced with the MOLGEP/MOLCAS codes, allowing
two-component GRECP/RCC-SD calculations in the intermediate coupling
scheme and one-component AGREP/NRCC-SDT calculations in the $\Lambda S$
coupling scheme. Nonrelativistic kinetic energy operators and
relativistic effective $j$-dependent potentials were employed in the
two-component calculations.

\section{Basis set}
\label{Basis}

The basis set for mercury was optimized in atomic RCC-SD calculations with the
help of the procedure proposed earlier\@.\cite{Mosyagin1}  The basis functions
were
generated in HF calculations of numerical orbitals for some neutral atomic or
positively charged ionic states. The HFJ code~\cite{Tupitsyn} was employed for
the HF calculations with the GRECP.

We start with a HF calculation of the $6s^2$ state of Hg with the GRECP
operator to obtain numerical $5s_{1/2},\allowbreak 5p_{1/2},\allowbreak
5p_{3/2},\allowbreak 5d_{3/2},\allowbreak 5d_{5/2}, \allowbreak 6s_{1/2}$
pseudospinors.  The $6p_{1/2}$ and $6p_{3/2}$ pseudospinors are then derived
from numerical calculations for the $LS$ averages of the $[6s^1] 6p^1$
configuration. The configuration notation used here puts in square
brackets shells frozen after the initial calculation of the $6s^2$ state;
the $5s,5p$ and $5d$ shells are understood to be in the square brackets and
are dropped for brevity.

The $\tilde{\varphi}_{5s},\allowbreak \tilde{\varphi}_{5p},\allowbreak
\tilde{\varphi}_{5d},\allowbreak \tilde{\varphi}_{6s},\allowbreak
\tilde{\varphi}_{6p}$ and $\Delta\tilde{\varphi}_{5p},\allowbreak
\Delta\tilde{\varphi}_{5d},\allowbreak \Delta\tilde{\varphi}_{6p}$
numerical radial orbitals are derived from the HF orbitals
$\tilde{\varphi}_{nl\pm}$, with $\pm$ denoting $j=l\pm1/2$, by
\begin{eqnarray}
\tilde{\varphi}_{ns}(r)  = & \tilde{\varphi}_{ns+}(r) , \\
\tilde{\varphi}_{nl}(r)  = & N[\tilde{\varphi}_{nl+}(r)+
                              \tilde{\varphi}_{nl-}(r)]
			                &  \mbox{~~~~~for~}  l=1,2,
\label{av} 			                                \\
\Delta\tilde{\varphi}_{nl}(r)   = & N'[\tilde{\varphi}_{nl+}(r)-
                              \tilde{\varphi}_{nl-}(r)]
			                & \mbox{~~~~~for~}   l=1,2,
\end{eqnarray}
where $N$ and $N'$ are normalization factors. The reference basis set is
constructed from the $5s,\allowbreak 5p,\allowbreak 5d,\allowbreak
6s,\allowbreak 6p$ orbitals. An RCC-SD calculation with the GRECP operator is
carried out in this basis with the 18 external electrons of Hg correlated.
The Hg$^+$ and Hg$^{2+}$ have, obviously, one or two fewer correlated electrons.
The Fock-space scheme for this calculation is

\begin{equation}
{\rm Hg}^{2+} \rightarrow {\rm Hg}^+ \rightarrow {\rm Hg} \;,
\end{equation}
with electrons added in the $6s$ and $6p$ one-electron states.

The next stage involves
HF calculations of a series of $7p$ orbitals with
the AGREP operator for configurations corresponding to the neutral Hg and a
set of Hg$^{n+}$ ions, namely $[5d^{10} 6s^1] 7p^1$, $[5d^{10-(n-1)}] 7p^1$
($n=1,2,\ldots 10$), $[5p^{6-(n-11)}] 7p^1$ ($n=11,12,\ldots 16$), etc. The
frozen $5p$ and $5d$ orbitals in these calculations are obtained from
Eq.~(\ref{av}).
The $7p$ orbitals derived this way
are localized in different spatial regions. A series of Schmidt-orthogonalized
basis sets is now formed by adding each of these $7p$ orbitals to the reference
basis. Basis sets obtained by addition of the
$\Delta\tilde{\varphi}_{5p}$ and $\Delta\tilde{\varphi}_{6p}$ orbitals are
also included. An RCC-SD
calculation of nine low-lying states (the ground $6s^2$ $^1\!S_0$, excited
$6s^16p^1$ $^3\!P_{0,1,2}$ and $^1\!P_1$ states of the neutral atom, $6s^1$
$^2\!S_{1/2}$ and $6p^1$ $^2\!P_{1/2,3/2}$ of Hg$^+$, and $^1\!S_0$ of
Hg$^{2+}$) with the GRECP operator is performed for each of these bases.
Similar series of
calculations are carried out for the $7s$ orbitals instead of the $7p$, and
also for the $6d,5f,5g$ orbitals. The principal quantum number of these
virtual orbitals is taken to be one higher than the maximum principal quantum
number of the corresponding orbitals in the reference basis set, to
avoid large overlap of the new and previously selected orbitals. For each
basis set, the largest change among all possible one-electron transition
(excitation or ionization) energies between the nine states listed
above relative to the results of the reference basis set is found.  This
change is then multiplied by a factor of $1/(2l+1)$, where $l$ is the angular
momentum quantum number of the added orbital. The orbital
which gives the largest energy change is added to the reference basis set.
The procedure is repeated for the next series of
virtual orbitals, resulting in a step by step expansion of the reference basis
and diminution of changes in the transition energies. The procedure is
terminated when the largest transition energy change after adding the orbital
goes down to about 15~cm$^{-1}$. The numerical radial orbitals are then
approximated by Gaussian functions and a $(14,12,9,3,2)/[7,7,4,2,1]$
basis set for molecular calculations is finally produced.
This algorithm of basis set generation is designed to account primarily
for correlation and spin-orbit effects which have different contributions
to the states under consideration, so that possible omissions in the resulting
basis will cause nearly state-independent errors and give accurate
transition energies.

It is important to note that the average radii and space localization
of the $f$ and $g$ correlation functions generated above are intermediate
between those of the $5d$ and $6s,6p$ functions. These $f$ and $g$ functions
describe primarily correlation between valence and outer core ($5d$) electrons.
The transition energy changes (without the $1/(2l+1)$ factor)
resulting from the addition of the $5f$, $5g$,
and $6f$ functions are up to 5800, 790, and 240~cm$^{-1}$, respectively.

The $(8,4,3)/[4,2,1]$ basis set from the ANO-L library~\cite{MOLCAS} is used
for hydrogen.

\section{Results and discussion}

As demonstrated in an earlier publication,\cite{Mosyagin1} the energy
contributions from correlations with the $5s$ and $4f$ shells of Hg largely
cancel each other.  These electrons may therefore be frozen in correlation
calculations, with resulting errors of up to 200~cm$^{-1}$ in one-electron
excitation or ionization energies. The molecular RCC-SD calculations in the
present paper are carried out with 19 correlated electrons of HgH (18 for
HgH$^+$).  The molecular orbital originating from the $5s$ orbital of Hg is
frozen after the HF calculation, while the $4f,4d,4p,4s$ and deeper core
electrons of Hg are excluded from explicit consideration by using the
20e-GRECP.

The computational effort in RCC-SD calculations increases rapidly with the
size of the basis. The highest orbital angular momentum in the basis was
therefore set to 4 ($g$-type harmonics).
This truncation leads to errors of up to 400~cm$^{-1}$ in one-electron
transition (excitation and ionization)
energies of the Hg atom (Ref.\ \citen{Mosyagin1}).
However, errors for transitions which do not
involve excitation or ionization of the tightly bound $6s$ electron
are below 150~cm$^{-1}$.
The inherent 20e-GRECP errors are up to 100~cm$^{-1}$ (Ref.\ \citen{Mosyagin1}).
Errors caused by using intermediate rather than $jj$ coupling are up to
100~cm$^{-1}$, as one can deduce from Table~\ref{Atom}, which reports
results of the 20e-GRECP/RCC-SD calculation on the Hg atom with the
18 external electrons correlated in the
$(14,12,9,3,2)/[7,7,4,2,1]$ basis described above.
The Fock-space scheme employed is
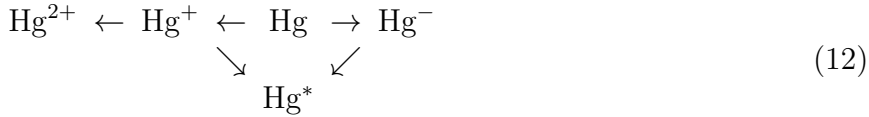
\begin{equation}
\begin{array}{ccccccc}
   {\rm Hg}^{2+} & \leftarrow &
    {\rm Hg}^+   & \leftarrow & {\rm Hg} & \rightarrow & {\rm Hg}^-      \\
 & &             & \searrow   &          & \swarrow    &                 \\
 & &             &            &{\rm Hg^*}&             &
\end{array}
\label{FS}
\end{equation}
with electrons added in the $6p$, $7s$, $7p$ orbitals and
removed from the $6s$ orbital.

In a previous paper,\cite{Mosyagin1} describing atomic Hg
calculations, it was concluded that neglect of virtual triple excitations
is the main cause for the differences between
experimental data and results of a 34-electron RCC-SD calculations with
a large basis including up to $h$-type functions ($l\leq 5$).  The contribution
of triples is estimated in the present work by resorting to
 spin-averaged (i.e., with AGREP only)
calculations and assuming that triples correction to the shells
correlated does not change much upon going to the fully
relativistic scheme. These
corrections are taken as the differences between the AGREP/NRCC-SDT and
AGREP/NRCC-SD total energies for each state. The $[7,7,4,2,1]$ Hg basis
generated above was used, and two cases were considered, with 12 and 18
electrons correlated (12e-T and 18e-T) in the Hg atom.
Table~\ref{Atom} shows that differences between the RCC-SD(12e-T) and
RCC-SD(18e-T) results are relatively small.  It is therefore sufficient to
take into account in molecular calculations triple cluster excitations of
the 12 external electrons of Hg.
The approximate inclusion of triple excitations for 18 electrons
in the intermediate coupling RCC-SD calculations
leads to an average error of 270~cm$^{-1}$ relative to experiment in the six
transitions listed in Table~\ref{Atom}. The largest error is 540~cm$^{-1}$, for
the excitation to the $^1{P}_1$ state.
Considering the Fock-space scheme~(\ref{FS}) and the approximations employed,
this is in line with results expected from the RCC-SDT method.

Molecular GRECP/RCC-SD and AGREP/NRCC-SDT calculations were carried out for
13 internuclear distances from 2.637 to 3.837~a.u., with intervals of
0.1~a.u.  Molecular spectroscopic constants were calculated
by the Dunham method in the Born-Oppenheimer approximation, using
the DUNHAM-SPECTR code.\cite{Mitin}

The effect of basis set superposition errors (BSSEs) in the 19-electron
RCC-SD calculation is demonstrated in Table~\ref{BSSE}. These errors are quite
sizable, up to 1800~cm$^{-1}$ for dissociation energies.
The main contribution to the BSSE arises from correlation in the core
region.
Details of the BSSE
calculation may be found in Ref.~\citen{BSSE}.
These errors are mainly due to limitations on the size of the basis which can
be used practically in molecular RCC calculation.
As stated in section~\ref{Basis},
the algorithm for generating the basis set aims primarily to optimize
excitation and ionization energies of the lowest lying states in mercury.
Since the transitions considered involve changes in valence shell occupation
only, the resulting basis may be nearly complete in the valence region but
deficient in the core region. The BSSE will therefore have weak dependence on
the valence-shell configuration, and may be estimated quite accurately
by the counterpoise correction (CPC) method. Table~\ref{BSSE} shows that
different valence states used to calculate the CPC yield very similar
spectroscopic constants.
The basis generated above is therefore compact enough and can be used in
precise calculations provided the CPC correction is applied.
All the GRECP/RCC-SD and AGREP/NRCC-SDT results reported in subsequent tables
were corrected using CPCs calculated for the Hg $6s^2$ state with a ghost H
atom.  CPCs calculated for the ground state of the H atom are about
1~cm$^{-1}$, and are therefore ignored.

Results for the ground states of HgH and HgH$^+$ are collected in
Table~\ref{HgH}. RCC-SD spectroscopic constants have sizable errors;
they also show considerable differences between the two Fock-space schemes.
Such differences are caused by the truncation of the CC cluster operator,
and indicate significant contributions of higher excitations.
The effect of virtual triple excitations to individual state energies was
estimated from the difference of AGREP/NRCC-SDT and AGREP/NRCC-SD values.
Two sets of triples calculations were carried out, correlating 3 and 13
electrons, and denoted by (3e-T) and (13e-T), respectively.
The full RCC-SD basis was used for 3 electrons, but a reduced basis,
[5,5,3,1] on Hg and [3,2] on H, was used in the 13-electron case.
Spectroscopic constants of the HgH ground state which include 13-electron
triples corrections are
in very good agreement with experiment. Dependence on the Fock-space scheme
is also minimal. Similar behavior is exhibited by the ground state of HgH$^+$.
Here, results obtained with scheme (\ref{FS1}) (i.e.\ in the $(0,0)$ Fock-space
sector) are slightly better that those of scheme (\ref{FS2}) (in the
$(2,0)$ sector) as might be expected. Still, differences are rather small.

Results for the excited $^2\Pi_{1/2}$ and $^2\Pi_{3/2}$ states with the
leading $\sigma^2 \pi^1$ configuration are presented in table~\ref{HgH_exc}.
The triples corrections to the spectroscopic constants are smaller than for
the HgH ground state, so that the RCC-SD values are in good agreement with
experiment. Inclusion of triple excitations does produce some improvement.

\section{Summary and conclusion}
Generalized relativistic effective core potentials were used in the framework
of the Fock-space relativistic coupled-cluster method to calculate
spectroscopic constants of HgH and HgH$^+$ ground states as well as HgH low
excited states.  Twenty one electrons were treated explicitly in the molecule,
of which 19 were correlated by RCC. Basis set superposition errors were
evaluated by counterpoise correction, and contributions of triple excitations
of the external 13 electrons were estimated (without accounting for spin-orbit
coupling)
in a reduced basis. The resulting spectroscopic constants of the HgH and
HgH$^+$ states are in very good agreement with experiment, with errors on the
order of a few mbohr for $R_e$, tens of wavenumbers for excitation energies
and vibrational frequencies, and proportionately for other properties.

\acknowledgments{
This work was supported by INTAS grant No 96--1266.
A~T and N~M thank the
Russian Foundation for Basic Research (grants No 99--03--33249 and
01--03--06335).
Work at TAU was supported by the Israel Science Foundation and by the
U.S.-Israel Binational Science Foundation. Calculations were mostly carried out
at TAU.
N~M thanks J~H~van Lenthe
for remarks made at the recent Conference
on Electron Correlations in a Relativistic Framework.
 The authors are grateful to A~B~Alekseyev for discussions.}

\begin{table}
\caption{ Ionization and excitation energies of the lowest-lying states of
   atomic mercury obtained from the 20e-GRECP/18e-RCC-SD calculations with
	  a $(14,12,9,3,2)/[7,7,4,2,1]$ basis set in the intermediate
          and $jj$- coupling schemes, without and with triples correction
          for 12 and 18 external electrons.
          All values are in cm$^{-1}$. }
\label{Atom}
\begin{tabular}{@{}lccccc}
 State          & Expt.$^{\rm a}$& RCC-SD   & RCC-SD & RCC-SD(12e-T) & RCC-SD(18e-T) \\
                       &              &\multicolumn{4}{c}{Coupling scheme}\\
\cline{3-6}
                       &              & $jj$     & intermediate & intermediate & intermediate \\
\hline
 \multicolumn{6}{l}{$5d^{10} 6s^2 (^1S_0) \rightarrow$}         \\
 $5d^{10} 6s^1 6p^1 (^3P_0)$ &  37645 &  37959   &  38055  &  37926 &  37971 \\
 $5d^{10} 6s^1 6p^1 (^3P_1)$ &  39412 &  39756   &  39826  &  39697 &  39742 \\
 $5d^{10} 6s^1 6p^1 (^3P_2)$ &  44043 &  44346   &  44322  &  44193 &  44238 \\
 $5d^{10} 6s^1 6p^1 (^1P_1)$ &  54069 &  54915   &  54914  &  54619 &  54607 \\
 $5d^{10} 6s^1 (^2S_{1/2})$  &  84184 &  83688   &  83680  &  84015 &  84240 \\
 \multicolumn{6}{l}{$5d^{10} 6s^1 (^2S_{1/2}) \rightarrow$}     \\
 $5d^{10} (^1S_0)$           & 151280 & 150018   & 149977  & 150898 & 151094 \\
\end{tabular}

\noindent $^{\rm a}$Ref.~\citen{Moore}.

\end{table}

\clearpage
\begin{table}
\caption{ Spectroscopic constants of
the ground states of
the HgH molecule and HgH$^+$ ion
          from GRECP/RCC-SD-1 calculations with different BSSE corrections.
$R_e$ is in \AA, $D_e$ in eV, $Y_{02}$ in $10^{-6}$~cm$^{-1}$,
other values in cm$^{-1}$.}
\label{BSSE}
\begin{tabular}{@{}lccccccc}
                            &$R_e$&$w_e$&$D_e$
&$B_e$&$w_e x_e$&$\alpha_e$&$-Y_{02}$\\
\hline
\multicolumn{8}{c}{HgH ($\sigma^2\sigma^1$) $^2\Sigma_{1/2}^+$}\\
  Uncorrected for BSSE         &  1.675  &  1686           &  0.50           & 5.99 & 57 & 0.262 & 306 \\
  CPC from Hg $6s^2$           &  1.702  &  1597           &  0.34           & 5.80 & 56 & 0.259 & 310 \\
  CPC from Hg $6s^1 6p^1$ $(^3P_0)$ &  1.701  &  1598           &  0.34           & 5.81 & 56 & 0.261 & 311 \\
  CPC from Hg $6s^1 6p^1$ $(^3P_1)$ &  1.701  &  1599           &  0.34           & 5.81 & 56 & 0.259 & 310 \\
  CPC from Hg $6s^1 6p^1$ $(^3P_2)$ &  1.701  &  1599           &  0.34           & 5.81 & 57 & 0.259 & 310 \\
  CPC from Hg $6s^1 6p^1$ $(^1P_1)$ &  1.701  &  1599           &  0.34           & 5.81 & 56 & 0.259 & 310 \\
\hline
\multicolumn{8}{c}{ HgH$^+$ ($\sigma^2$) $^1\Sigma_0^+$}\\
  Uncorrected for BSSE         &  1.570  &  2145           &  2.91           & 6.82 & 41 & 0.205 & 277 \\
  CPC from Hg $6s^2$           &  1.588  &  2067           &  2.72           & 6.66 & 39 & 0.199 & 278 \\
  CPC from Hg$^+$ $6s^1$       &  1.588  &  2069           &  2.73           & 6.67 & 39 & 0.199 & 278 \\
\end{tabular}

\end{table}

\begin{table}
\caption{ Spectroscopic constants of the ground states of
the HgH molecule and HgH$^+$ ion.
The GRECP/RCC results are corrected by CPCs
calculated for the Hg $6s^2$ state.
$R_e$ is in \AA, $D_e$ in eV, $Y_{02}$ in $10^{-6}$~cm$^{-1}$,
other values in cm$^{-1}$.}
\label{HgH}
\begin{tabular}{@{}lccccccc}
                            &$R_e$&$w_e$&$D_e$
&$B_e$&$w_e x_e$&$\alpha_e$&$-Y_{02}$\\
\hline
\multicolumn{8}{c}{HgH ($\sigma^2\sigma^1$) $^2\Sigma_{1/2}^+$}\\
\underline{Experiment}             &           &                 &                 &                   &                  &                  &                \\
 Ref.\ \citen{Herzberg}  & [1.766]$^{\rm a}$
                                        & [1203]$^{\rm a}$& ~0.46           & [5.39]$^{\rm a}$  &                  &                  & [395]$^{\rm a}$ \\
 Ref.\ \citen{Dufayard}  & ~1.741    & ~1385           & ~0.46           & ~5.55             & ~75              &  0.271           &                 \\
 Ref.\ \citen{Herzberg1} & ~1.740    & ~1387           & ~0.46           & ~5.55             & ~83              &  0.312           & ~344            \\
 Ref.\ \citen{Stwalley}  & ~1.735    & ~1421           & ~0.46           & ~5.59             & 121              &  0.404           & ~346            \\
\underline{Present calculations}   &           &                 &                 &                   &                  &                  &                \\
 GRECP/RCC-SD-1             & ~1.702    & ~1597           & ~0.34           & ~5.80             & ~56              &  0.259           & ~310            \\
 GRECP/RCC-SD-2             & ~1.730    & ~1419           & ~0.32           & ~5.61             & ~85              &  0.349           & ~361            \\
 GRECP/RCC-SD(3e-T)-1       & ~1.714    & ~1528           & ~0.40           & ~5.72             & ~65              &  0.287           & ~326            \\
 GRECP/RCC-SD(3e-T)-2       & ~1.733    & ~1386           & ~0.37           & ~5.60             & ~92              &  0.374           & ~376            \\
GRECP/RCC-SD(13e-T)-1      &  ~1.730    &  ~1424           &  ~0.41            &  ~5.62             &  ~81              &   0.343           &  ~358            \\
GRECP/RCC-SD(13e-T)-2      &  ~1.738    &  ~1362           &  ~0.41            &  ~5.56             &  ~93              &   0.380           &  ~382            \\
\underline{Other calculations}     &           &                 &                 &                   &                  &                  &                \\
 RECP/MRD-CI\cite{Alekseyev}& ~1.777    & ~1309           & ~0.32           &                   &                  &                  &                 \\
 PP/ACPF+SO\cite{Hausser}$^{\rm ,b}$
                            & ~1.722    & ~1414           & ~0.44           &                   &                  &                  &                 \\
 RESC/CCSD+CPC\cite{Motegi} & ~1.763    & ~1263           & ~0.28           &                   &                  &                  &                 \\
 AIMP/CCSD+CPC\cite{Motegi} & ~1.762    & ~1264           & ~0.29           &                   &                  &                  &                 \\
\hline
\multicolumn{8}{c}{HgH$^+$ ($\sigma^2$) $^1\Sigma_0^+$}\\
\underline{Experiment}             &           &                 &                 &                   &                  &                  &                \\
 Ref.\ \citen{Herzberg}  & ~1.594    & ~2028           & (3.11)$^{\rm c}$& ~6.61 & ~41   & 0.206 & ~285  \\
 Ref.\ \citen{Herzberg1} & ~1.594    & ~2034           & (2.4)$^{\rm c}$ & ~6.61 & ~46   & 0.206 & ~285  \\
\underline{Present calculations}   &           &                 &                 &                   &                  &                  &                \\
 GRECP/RCC-SD-1$^{\rm d}$   & ~1.588    & ~2067           & ~2.72           & ~6.66 & ~39   & 0.199 & ~278  \\
 GRECP/RCC-SD-2             & ~1.586    & ~2149           & ~2.42           & ~6.68 & ~21   & 0.153 & ~259  \\
 GRECP/RCC-SD(3e-T)-2       & ~1.592    & ~2060           & ~2.55           & ~6.63 & ~31   & 0.187 & ~276  \\
GRECP/RCC-SD(13e-T)-1      &  ~1.591     &  ~2044            &  ~2.70           &  ~6.64  &  ~41    &  0.208 &  ~282  \\
GRECP/RCC-SD(13e-T)-2      &  ~1.596     &  ~2051            &  ~2.59           &  ~6.60  &  ~32    &  0.185 &  ~275  \\
\underline{Other calculations}     &           &                 &                 &                   &                  &                  &                \\
 PP/ACPF\cite{Hausser}$^{\rm ,b}$
                            & ~1.593    & ~1959           & ~2.69           &       &       &       &      \\
\end{tabular}

\noindent $^{\rm a}$Cited in Ref.~\citen{Herzberg}
                    as corresponding to the zero vibrational level.

\noindent $^{\rm b}$See original work~\cite{Hausser} for results
                    derived by other calculation methods.

\noindent $^{\rm c}$Cited in Ref.~\citen{Herzberg,Herzberg1}
                    as uncertain.

\noindent $^{\rm d}$GRECP/RCC-SD(3e-T)-1 results are identical with
                    GRECP/RCC-SD-1, as triple
		    excitations do not occur for two
		    correlated electrons.

\end{table}

\begin{table}
\caption{Spectroscopic constants for excited states of
the HgH molecule.
The GRECP/RCC results are corrected by CPCs
calculated for the Hg $6s^2$ state.
$R_e$ is in \AA, $Y_{02}$ in $10^{-6}$~cm$^{-1}$,
other values in cm$^{-1}$.}
\label{HgH_exc}
\begin{tabular}{@{}lccccccc}
                            &$R_e$&$w_e$&$T_e$
&$B_e$&$w_e x_e$&$\alpha_e$&$-Y_{02}$\\
\hline
\multicolumn{8}{c}{HgH$^*$ ($\sigma^2\pi^1$) $^2\Pi_{1/2}$}\\
\underline{Experiment}             &           &                 &                 &                   &                  &                  &                \\
 Ref.\ \citen{Herzberg}  & [1.601]$^{\rm a}$
                                        &[1939]$^{\rm a}$ &                 & [6.56]$^{\rm a}$  &                  &                  & [285]$^{\rm a}$ \\
 Ref.\ \citen{Herzberg1} & ~1.586    & ~2066           & 24578           & ~6.68             & [64]$^{\rm a}$   & [0.242]$^{\rm a}$&                 \\
 Ref.\ \citen{Dufayard}  & ~1.583    & ~2068           & 24590           & ~6.70             & ~65              & ~0.267           &                 \\
 Ref.\ \citen{Mayama}    & ~1.583    & ~2031           & 24609           & ~6.71             & ~47              & ~0.219           &                 \\
\underline{Present calculations}   &           &                 &                 &                   &                  &                  &                \\
 GRECP/RCC-SD-1             & ~1.578    & ~2100           & 24044           & ~6.75             & ~39              & ~0.201           & ~280            \\
 GRECP/RCC-SD(3e-T)-1       & ~1.581    & ~2080           & 24229           & ~6.72             & ~40              & ~0.205           & ~283            \\
GRECP/RCC-SD(13e-T)-1      &  ~1.582    &  ~2065           &  24688           &  ~6.71             &  ~44              &  ~0.215           &  ~286            \\
\underline{Other calculations}     &           &                 &                 &                   &                  &                  &                \\
 PP/CASSCF+MRCI+CIPSO\cite{Hausser}
                            & ~1.603    & ~1946           & 25004           &                   &                  &                  &                 \\
 RECP/MRD-CI\cite{Alekseyev}& ~1.615    & ~2023           & 25664           &                   &                  &                  &                 \\
\hline
\multicolumn{8}{c}{HgH$^*$ ($\sigma^2\pi^1$) $^2\Pi_{3/2}$}\\
\underline{Experiment}             &           &                 &                 &                   &                  &                  &                \\
 Ref.\ \citen{Dufayard}  & ~1.581    & ~2091           & 28283           & ~6.73             & ~61              & ~0.200           &                 \\
 Ref.\ \citen{Herzberg}  & ~1.579    & ~2068           & 28274           & ~6.74             & ~43              & ~0.230           & [282]$^{\rm a}$ \\
 Ref.\ \citen{Herzberg1} & ~1.580    & ~2067           & 28256           & ~6.73             & ~42              & ~0.214           &                 \\
\underline{Present calculations}   &           &                 &                 &                   &                  &                  &                \\
 GRECP/RCC-SD-1             & ~1.576    & ~2117           & 27629           & ~6.77             & ~37              & ~0.197           & ~278            \\
 GRECP/RCC-SD(3e-T)-1       & ~1.578    & ~2098           & 27815           & ~6.75             & ~38              & ~0.200           & ~281            \\
GRECP/RCC-SD(13e-T)-1   &~1.579   &~2083          &28275          &~6.74            &~42             &~0.210          &~284           \\
\underline{Other calculations}     &           &                 &                 &                   &                  &                  &                \\
 PP/CASSCF+MRCI+CIPSO\cite{Hausser}
                            & ~1.610    & ~1930           & 28714           &                   &                  &                  &                 \\
 RECP/MRD-CI\cite{Alekseyev}& ~1.615    & ~2033           & 28490           &                   &                  &                  &                 \\
\end{tabular}

\noindent $^{\rm a}$Cited in Ref.~\citen{Herzberg,Herzberg1}
                    as corresponding to the zero vibrational level.

\end{table}


\begin{thebibliography}{99}
\bibitem{Porter}        T.~L.~Porter, J.\ Opt.\ Soc.\ Am.\ {\bf 52}, 1201 (1962);
                        L.~B.~Jr.~Knight and W.~Jr.~Weltner, J.\ Chem.\ Phys.\ {\bf 55}, 2061 (1971).
\bibitem{Stwalley}      W.~C.~Stwalley, J.\ Chem.\ Phys.\ {\bf 63}, 3062 (1975).
\bibitem{Mayama}        S.~Mayama, S.~Hiraoka and K.~Obi, J.\ Chem.\ Phys.\ {\bf 81}, 4760 (1984).
\bibitem{Dufayard}      J.~Dufayard, B.~Majournat and O.~Nedelec, Chem.\ Phys.\ {\bf 128}, 537 (1988).
\bibitem{Das}           G.~Das and A.~C.~Wahl, J.\ Chem.\ Phys.\ {\bf 64}, 4672 (1976);
                        P.~J.~Hay, W.~R.~Wadt, L.~R.~Kahn and F.~W.~Bobrowicz, J.\ Chem.\ Phys.\ {\bf 69}, 984 (1978).
\bibitem{Hausser}       U.~H\"aussermann, M.~Dolg, H.~Stoll, H.~Preuss, P.~Schwerdtfeger and R.~M.~Pitzer,
                        Mol.\ Phys.\ {\bf 78}, 1211 (1993).
\bibitem{Alekseyev}     A.~B.~Alekseyev, H.-P.~Liebermann, R.~J.~Buenker and G.~Hirsch,
                        J.\ Chem.\ Phys.\ {\bf 104}, 4672 (1996).
\bibitem{Motegi}        K.\ Motegi, T.\ Nakajima, K.\ Hirao and L.\ Seijo,
                        J.\ Chem.\ Phys.\ {\bf 114}, 6000 (2001).
\bibitem{Tupitsyn}      I.~I.~Tupitsyn, N.~S.~Mosyagin and A.~V.~Titov, J.\ Chem.\ Phys.\ {\bf 103}, 6548 (1995).
\bibitem{Mosyagin}      N.~S.~Mosyagin, A.~V.~Titov and Z.~Latajka, Int.\ J.\ Quant.\ Chem.\ {\bf 63}, 1107 (1997).
\bibitem{Mosyagin1}     N.~S.~Mosyagin, E.~Eliav, A.~V.~Titov and U.~Kaldor, J.\ Phys.\ {\bf B 33}, 667 (2000).
\bibitem{Ross}          R.~B.~Ross, J.~M.~Powers, T.~Atashroo, W.~C.~Ermler, L.~A.~LaJohn and P.~A.~Christiansen,
                        J.\ Chem.\ Phys.\ {\bf 93}, 6654 (1990).
\bibitem{Titov}         A.~V.~Titov and N.~S.~Mosyagin, Int.\ J.\ Quant.\
                        Chem.\ {\bf 71}, 359 (1999);
                        A.~V.~Titov and N.~S.~Mosyagin, Rus.\ J.\ Phys.\ Chem.\
                       [Zh.\ Fiz.\ Khimii] {\bf 74} (Suppl.~2), S376 (2000),
                        E-print: http://xxx.lanl.gov/abs/physics/0008160 .
\bibitem{Titov1}        A.~V.~Titov, A.~O.~Mitrushenkov and I.~I.~Tupitsyn, Chem.\ Phys.\ Lett.\ {\bf 185}, 330 (1991).
\bibitem{Christ}        P.~A.~Christiansen, Y.~S.~Lee and K.~S.~Pitzer, J.\ Chem.\ Phys.\ {\bf 71}, 4445 (1979).
\bibitem{Comment1}      N.~S.~Mosyagin and A.~V.~Titov,
                        E-print: http://xxx.lanl.gov/abs/physics/9808006 (1998);
                        A.~V.~Titov and N.~S.~Mosyagin,
                        E-print: http://xxx.lanl.gov/abs/physics/0008239 (2000).
\bibitem{Eliav}         E.~Eliav, U.~Kaldor and Y.~Ishikawa, Phys.\ Rev.\ A {\bf 52}, 2765 (1995).
\bibitem{kaldorrevs}    U.~Kaldor, Few-Body Systems Suppl.\ {\bf 8}, 67 (1995);
                        Y.~Ishikawa and U.~Kaldor, in {\it Computational
                        Chemistry: Review of Current Trends}, edited by J.~Leszczynski
                        (World Scientific, Singapore, 1996), Vol.~1, p.~1;
                        U.~Kaldor, in {\it Recent Advances in Coupled-Cluster
                        Methods}, edited by R.~J.~Bartlett (World Scientific,
			Singapore, 1997) p.~125;
                        U.~Kaldor and E.~Eliav, Adv.\ Quantum Chem.\ {\bf 31}, 313 (1998).
\bibitem{MOLGEP}        A.~V.~Titov, A.~N.~Petrov, A.~I.~Panin and Yu.~G.~Khait,
                        {\it MOLGEP code for calculation of matrix elements
			with GRECP} (St.-Petersburg, 1999).
\bibitem{MOLCAS}        K.~Andersson, M.~R.~A.~Blomberg, M.~P.~F\"ulscher,
                        G.~Karlstr\"om, R.~Lindh, P.-A.~Malmqvist, P.~Neogr\'ady,
		        J.~Olsen, B.~O.~Roos, A.~J.~Sadlej, M.~Sch\"utz, L.~Seijo,
		        L.~Serrano-Andr\'es, P.~E.~M.~Siegbahn and P.-O.~Widmark,
                        {\it MOLCAS}, Version {\bf 4.1}
			(Lund University, Sweden, 1997).
\bibitem{Mitin}         A.~V.~Mitin, J.\ Comput.\ Chem.\ {\bf 19}, 94 (1998).
\bibitem{BSSE}          M.~Gutowski, J.~H.~Van Lenthe, J.~Verbeek,
                        F.~B.~Van Duijneveldt and G.~Chalasinski,
                        Chem.\ Phys.\ Lett.\ {\bf 124}, 370 (1986);
                        B.~Liu and A.~D.~McLean, J.\ Chem.\ Phys.\ {\bf 91}, 2348 (1989).
\bibitem{Moore}         C.~E.~Moore, Circ.\ Natl.\ Bur.\ Stand.\ (U.S.)
                        {\bf 467} (1958).
\bibitem{Herzberg}      K.~P.~Huber and G.~Herzberg,
                        {\it Molecular spectra and Molecular structure}. IV.
                        {\it Constants of Diatomic Molecules}
                        (Van Nostrand Reinhold, New York, 1979).
\bibitem{Herzberg1}     G.~Herzberg,
                        {\it Molecular spectra and Molecular structure}. I.
                        {\it Spectra of Diatomic Molecules}
                        (Van Nostrand Reinhold, New York, 1950).
\end{thebibliography}
\end{document}